\documentclass[prb,twocolumn,amsmath,amssymb,superscriptaddress]{revtex4-2}

\usepackage[pdftex]{graphics}
\usepackage{graphicx}
\usepackage{hyperref}
\usepackage{xcolor}
\usepackage{siunitx}
\usepackage{physics}
\usepackage{amsmath}
\usepackage{amssymb}

\begin{document}

\title{Spin-orbital-lattice entanglement in the ideal $j$\,=\,1/2 compound K$_2$IrCl$_6$}

\author{P. Warzanowski}
\author{M. Magnaterra}
\affiliation{Institute of Physics II, University of Cologne, 50937 Cologne, Germany}
\author{Ch. J. Sahle}
\affiliation{ESRF, The European Synchrotron, 71 Avenue des Martyrs, CS40220, 38043 Grenoble Cedex 9, France}
\author{M. Moretti Sala}
\affiliation{Dipartimento di Fisica, Politecnico di Milano, I-20133 Milano, Italy}
\author{P.~Becker}
\author{L.~Bohat\'{y}}
\affiliation{Sect. Crystallography, Institute of Geology and Mineralogy, University of Cologne, 50674 Cologne, Germany}
\author{I.~C\'{i}sa\v{r}ov\'{a}}
\affiliation{Department of Inorganic Chemistry, Charles University in Prague, 128 43 Prague 2, Czech Republic}
\author{G. Monaco}
\affiliation{Dipartimento di Fisica e Astronomia "Galileo Galilei", Universit\`{a} di Padova, I-35121 Padova, Italy}
\author{T. Lorenz}
\author{P.H.M. van Loosdrecht}
\affiliation{Institute of Physics II, University of Cologne, 50937 Cologne, Germany}
\author{J. van den Brink}
\affiliation{Institute for Theoretical Solid State Physics, IFW Dresden, 01069 Dresden, Germany}
\affiliation{Institute for Theoretical Physics and W\"urzburg-Dresden Cluster of Excellence ct.qmat, Technische Universit\"at Dresden, 01069 Dresden, Germany}
\author{M. Gr\"{u}ninger}
\affiliation{Institute of Physics II, University of Cologne, 50937 Cologne, Germany}

\begin{abstract}
Mott insulators with spin-orbit entangled $j$\,=\,1/2 moments host intriguing 
magnetic properties. The $j$\,=\,1/2 wave function requires cubic symmetry, 
while a noncubic crystal field mixes $j$\,=\,1/2 and 3/2 character. 
Spectroscopic studies of $5d^5$ iridates typically claim noncubic symmetry, 
e.g., based on a splitting of the excited $j$\,=\,3/2 quartet. 
A sizable splitting is particularly puzzling in antifluorite-type K$_2$IrCl$_6$, 
a frustrated fcc quantum magnet with global cubic symmetry. 
It raises the fundamental question about the stability of $j$\,=\,1/2 moments 
against magneto-elastic coupling. 
Combining resonant inelastic x-ray scattering with optical spectroscopy, 
we demonstrate that the multi-peak line shape in K$_2$IrCl$_6$ reflects a 
vibronic character of the $j$\,=\,3/2 states rather than a noncubic crystal field. 
The quasimolecular crystal structure with well separated IrCl$_6$ octahedra explains 
the existence of well-defined sidebands that are usually smeared out in solids.
Our results highlight the spin-orbital-lattice entangled character of cubic K$_2$IrCl$_6$ 
with ideal $j$\,=\,1/2 moments. 
\end{abstract}

\date{July 16, 2024}

\maketitle

\section{Introduction}

\begin{figure}[t]
	\centering
	\includegraphics[width=\columnwidth]{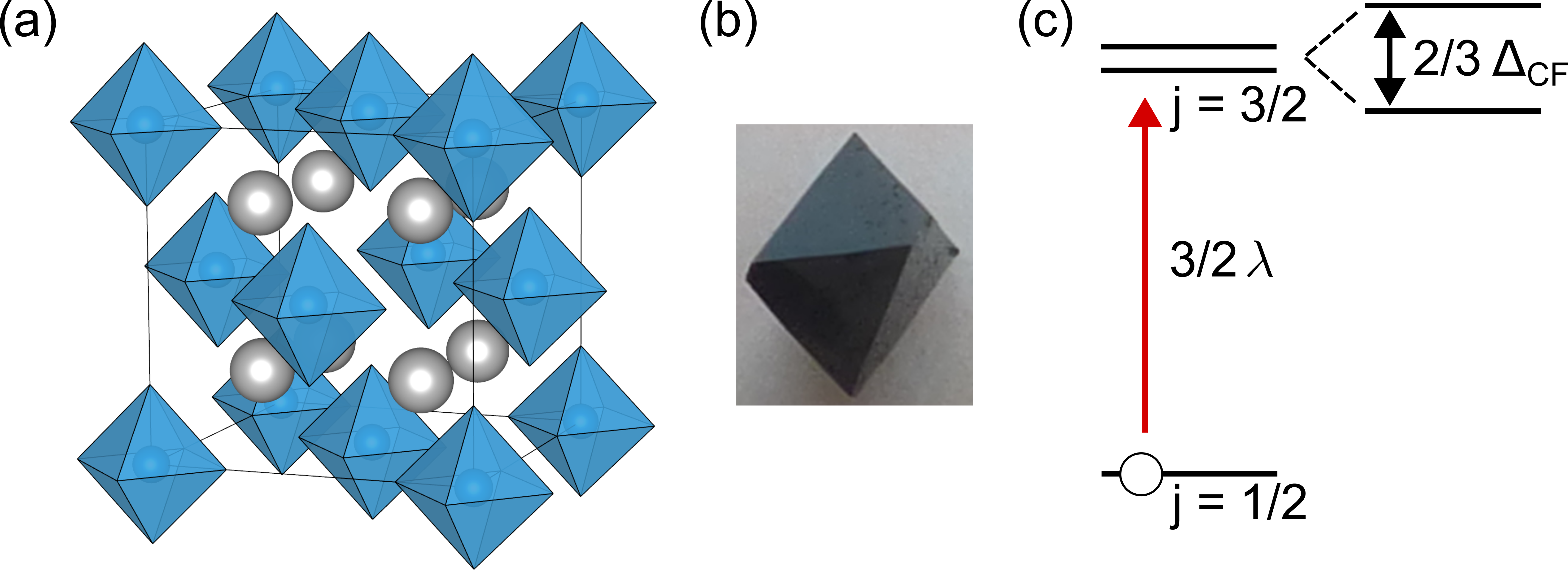}
	\caption{(a) Sketch of the cubic crystal structure of K$_2$IrCl$_6$ with well separated 
	IrCl$_6$ octahedra (blue). The K$^+$ ions are shown in gray.
    (b) Example of a grown single crystal of K$_2$IrCl$_6$.
	(c) The spin-orbit exciton denotes the excitation from the local $j$\,=\,1/2 ground state 
	to the $j$\,=\,3/2 excited state. In cubic symmetry, the energy equals 1.5\,$\lambda$. 
	The degeneracy of the quartet is lifted in a noncubic crystal field. 
	} 
	\label{fig:structure}
\end{figure}

The entanglement of spin and orbital degrees of freedom via strong spin-orbit coupling leads 
to a multitude of novel quantum magnetic phases in $4d$ and $5d$ transition-metal compounds
\cite{WitczakKrempa14,Rau16,Schaffer16,Takagi19,Streltsov20,Takayama21,Khomskii21}.
Particular interest has focused on the physics of $j$\,=\,1/2 moments in $4d^5$ Ru$^{3+}$ 
and $5d^5$ Ir$^{4+}$ compounds. Despite strong spin-orbit coupling, these allow for isotropic 
Heisenberg exchange in 180$^\circ$ corner-sharing bonding geometry but also yield Ising 
exchange in 90$^\circ$ edge-sharing configuration \cite{Jackeli09}, opening the door for 
the realization of Kitaev spin liquids with bond-directional exchange \cite{Chun15,Magnaterra23}
on tricoordinated lattices. 
Such local $j$\,=\,1/2 moments are formed by, e.g., $t_{2g}^5$ Ir$^{4+}$ ions in octahedral configuration. 
Resonant inelastic x-ray scattering (RIXS) at the Ir $L_3$ edge is a sensitive tool to test 
the $j$\,=\,1/2 character, probing excitations to the $j$\,=\,3/2 quartet, i.e., 
the spin-orbit exciton. A noncubic crystal-field contribution $\Delta_{\rm CF}$ lifts 
the degeneracy of the quartet, see Fig.\ \ref{fig:structure}(c), giving rise to an
admixture of $j$\,=\,3/2 character to the $j$\,=\,1/2 ground state 
wavefunction \cite{Moretti14CEF}. 
In Ir$^{4+}$ materials, RIXS typically detects such deviations from pure $j$\,=\,1/2 
character with crystal-field splittings of roughly 0.1\,eV 
\cite{Liu12,Gretarsson13,Rossi17,Aczel19,Revelli19,Ruiz21,delaTorre21,Jin22,Magnaterra23Ti,delaTorre23,Kim12,Kim14,Lu18,Kim12327,Moretti15}.

Surprisingly, a sizable splitting of the $j$\,=\,3/2 states has even been reported 
for compounds that are found to exhibit cubic symmetry in x-ray diffraction such as 
the double perovskite Ba$_2$CeIrO$_6$ and the antifluorite-type halides K$_2$Ir$X$$_6$ 
($X$\,=\,Cl, Br) \cite{Revelli19,Aczel19,Khan19,Reig20,Khan21,Bhaskaran21,Lee22,Meggle23,Bertin24}. 
The halides show such splitting in Ir $L$-edge RIXS, Raman spectroscopy, 
and infrared absorption \cite{Reig20,Khan21,Lee22,Meggle23}. 
All three compounds host the local moments on an fcc lattice, giving rise to highly frustrated 
quantum magnetism where a spin-liquid phase emerges in the phase diagram based on the 
geometric frustration of antiferromagnetic nearest-neighbor Heisenberg exchange 
augmented by next-nearest-neighbor exchange \cite{Revelli19,Balla20}.
Remarkably, finite antiferromagnetic Kitaev exchange in this case reduces the magnetic 
frustration and stabilizes long-range magnetic order \cite{Revelli19,Khan19,Khan21}. 
However, the pronounced frustration on the fcc lattice can be lifted via magneto-elastic 
coupling, where even small lattice distortions may cause strong variation of 
nearest-neighbor exchange couplings \cite{Revelli19}. 
Such small distortions may be reconciled with an apparent global cubic structure in 
diffraction experiments if distortions are essentially local and exhibit a very short 
correlation length. This raises the fundamental question on the stability of 
cubic $j$\,=\,1/2 moments in frustrated quantum magnets.

Recently, Iwahara and Furukawa \cite{Iwahara23} suggested an alternative scenario 
for K$_2$IrCl$_6$ in which the two-peak structure of the spin-orbit exciton observed 
in RIXS is attributed to the interplay of spin-orbit coupling and 
electron-phonon coupling that gives rise to vibronic sidebands, 
i.e., spin-orbital-lattice entangled states with a mixed vibrational-electronic character. 
This scenario of a dynamic Jahn-Teller effect in the $j$\,=\,3/2 \textit{excited} state 
\cite{Plotnikova16,Streltsov20} does not require a local breaking of cubic symmetry in 
the ground state and hence reconciles the spectroscopic data with the cubic symmetry 
reported in diffraction \cite{Khan19,Reig20,Bertin24} and in electron spin resonance, 
where an isotropic $g$ factor is found \cite{Bhaskaran21}. 
However, this explanation raises the puzzling question why distinct vibronic sidebands 
thus far have not been reported in the compelling series of experimental $L$-edge RIXS 
studies of other $5d^5$ Ir$^{4+}$ oxides and halides
\cite{Liu12,Gretarsson13,Rossi17,Aczel19,Revelli19,Ruiz21,delaTorre21,Jin22,Magnaterra23Ti,delaTorre23,Kim12,Kim14,Kim12327,Moretti15,Moretti14}. 
Based on the excellent energy resolution of optical spectroscopy, vibronic sidebands 
of the spin-orbit exciton have been observed in optical data of the $4d^5$ $j$\,=\,1/2 
Kitaev material $\alpha$-RuCl$_3$ \cite{Warzanowski20,Lee21}.
Furthermore, a dressing of the spin-orbit exciton with phonon sidebands has been 
claimed in oxygen $K$-edge RIXS on the related Kitaev material $\alpha$-Li$_2$IrO$_3$ 
\cite{Vale19}. 
Compared to $L$-edge RIXS, however, RIXS at the O $K$ edge is much more sensitive to 
vibrational features due to the very different character of the intermediate state 
in the scattering process and the much longer core hole life time. 
This has recently been demonstrated for $5d^1$ Ba$_2$NaOsO$_6$ \cite{Agrestini24} 
and is further exemplified by the observation of an entire ladder of strong phononic 
peaks in $\alpha$-Li$_2$IrO$_3$ at the O $K$ edge \cite{Vale19} and the absence of 
any phononic features in $L$-edge RIXS of the same compound \cite{Revelli20}. 
Beyond iridates, the observation of vibronic sidebands of electronic excitations 
by means of transition-metal $L$-edge RIXS has been claimed in
$3d^9$ Ca$_2$Y$_2$Cu$_5$O$_{10}$ \cite{Lee14}, 
$4d^4$ K$_2$RuCl$_6$ \cite{Iwahara23Ru,Takahashi21}, 
and $5d^1$ $A_2$MgReO$_6$ ($A$\,=\,Ca, Sr, Ba) \cite{Frontini24} 
but the observed features are broad and individual sidebands are not or hardly resolved. 
This is the typical situation in solids. 
While molecules like O$_2$ exhibit distinct vibronic sidebands 
of electronic excitations \cite{Hennies10}, in solids the existence of many different 
phonon modes and their dispersion smear out the sideband structure,
most often turning the line shape into a broad hump even in optical data 
\cite{Figgis,Henderson89,Ballhausen,Rueckamp05}. 
Remarkably, individual vibronic sidebands have been resolved in optical data 
of quasimolecular crystals such as K$_3$NiO$_2$ with isolated NiO$_2$ units \cite{Figgis}, 
and the same characteristic features have been detected recently in optical data of the 
sister compounds K$_2$ReCl$_6$ and K$_2$OsCl$_6$ \cite{Warzanowski23,Warzanowski24}.

Here, we join forces of RIXS and optical spectroscopy to thoroughly study the 
spin-orbit exciton in K$_2$IrCl$_6$. Phenomenologically, the two-peak structure seen
in RIXS can be explained by either a noncubic crystal field or a vibronic picture. 
In contrast, the excellent energy resolution of the optical data allows us to resolve 
a multi-peak structure that highlights the vibronic origin, respecting cubic symmetry. 
These excitations are particularly well defined in K$_2$IrCl$_6$ due to its quasimolecular 
crystal structure with spatially isolated IrCl$_6$ octahedra. 
The competition of spin-orbit coupling and electron-phonon coupling 
gives rise to a dynamic Jahn-Teller effect in the $j$\,=\,3/2 excited state, 
hybridizing the spin-orbit exciton with vibrational states \cite{Iwahara23}. 
Empirically, the overall line shape of these spin-orbital-lattice entangled excitations 
can be described by the Franck-Condon approximation, where the eigenstates are product states 
of electronic and vibrational states. However, the spin-orbital-lattice entangled nature 
is evident in the parameters, 
in particular in the peak splitting and its temperature dependence. 
We further demonstrate that the contribution of elementary phonon excitations in 
Ir $L_3$-edge RIXS is negligible in K$_2$IrCl$_6$. 
In contrast, the vibronic "phonon" sidebands of the spin-orbit exciton 
contribute in a direct RIXS process and are resonantly enhanced due to the electronic 
part of the wave function. 
Finally, we observe the double spin-orbit exciton at 1.3\,eV in $\sigma_1(\omega)$ 
which also shows vibronic sidebands. 
Our results firmly establish the spin-orbital-lattice entangled nature of 
$j$\,=\,3/2 excited states in cubic K$_2$IrCl$_6$ with ideal $j$\,=\,1/2 moments.

\section{Experimental}
\label{sec:spectroscopy}

\begin{figure}[t]
	\centering
	\includegraphics[width=\columnwidth]{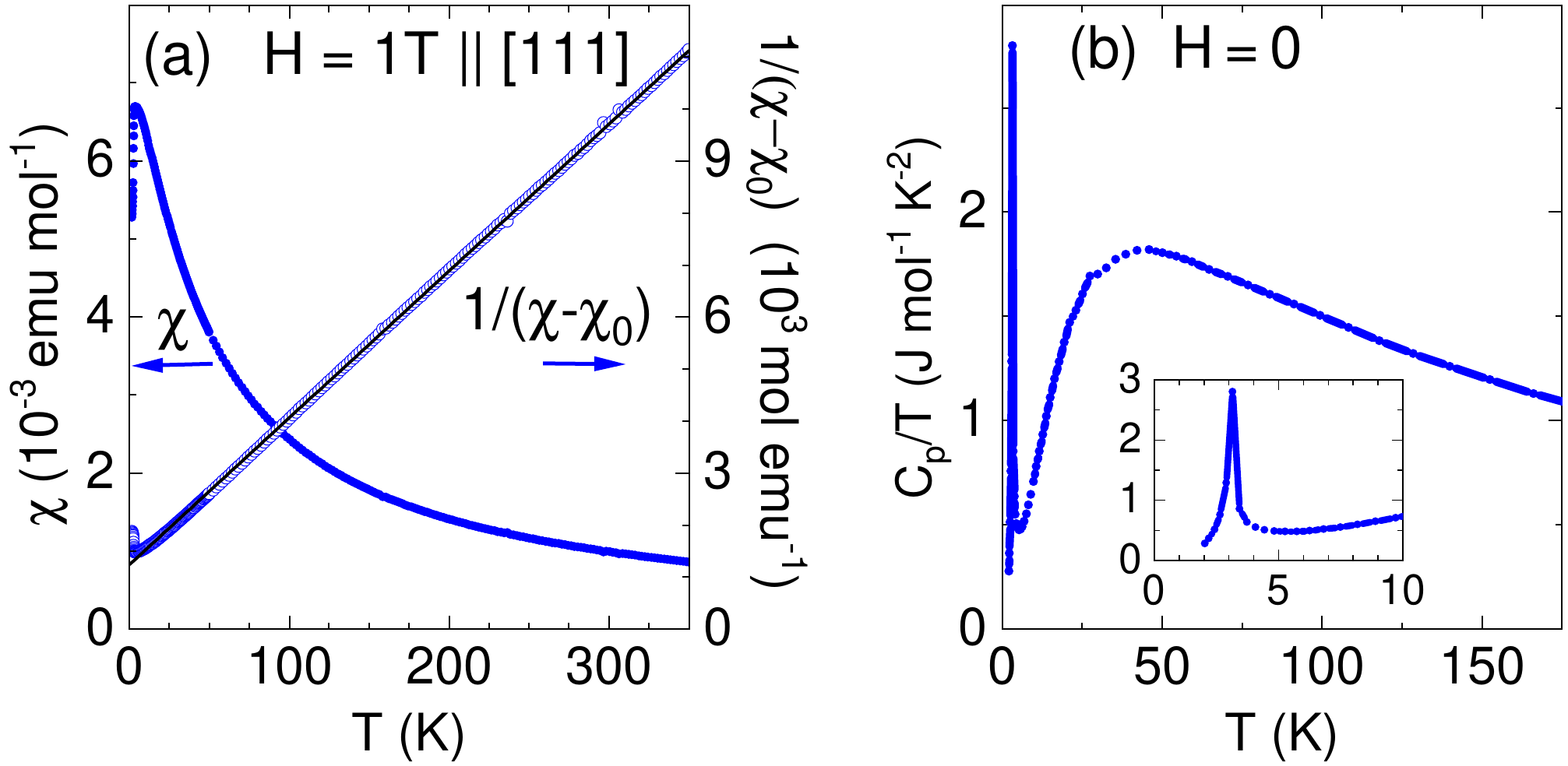}
	\caption{(a) The magnetic susceptibility $\chi$ of K$_2$IrCl$_6$ exhibits 
		Curie-Weiss behavior as seen by the straight line of $1/(\chi-\chi_0)$ (right axis) 
		with $\chi_0$\,=\,$-3.9 \times 10^{-5}$emu/mol representing the sum of 
		core diamagnetism and Van Vleck paramagnetism. The linear fit (black) above 
		100\,K yields a Weiss temperature $\theta_{\rm W}$\,=\,-43.9\,K and an effective magnetic moment 
		$\mu_{\rm eff}$\,=\,1.69\,$\mu_{\rm B}$. 
		(b) The specific heat shows a pronounced peak signaling antiferromagnetic order 
		at  $T_{\rm N}$\,=\,3.1\,K, as plotted on an enlarged scale 
		in the inset.}
	\label{fig:cp_chi}
\end{figure}

High-quality single crystals were grown from a solution of 
commercially available K$_2$IrCl$_6$ in $\approx$\,5.2\,molar HCl by controlled evaporation 
of the solvent at 293\,K.\@ Within typical growth periods of two weeks, crystals of dimensions 
up to 1$\times$1$\times$2\,mm$^{3}$ were obtained, see Fig.\ \ref{fig:structure}(b). 
In single-crystal x-ray diffraction measurements performed at ten different temperatures 
from 120 to 290\,K, we find the cubic space group $Fm\bar{3}m$ (Nr.\ 225) and lattice constants 
$a$\,=\,9.7458(2)\,\AA{} (290\,K) and 9.6867(3)\,\AA{} (120\,K), 
see Appendix A.\@ Our x-ray diffraction results agree very well 
with earlier data measured at 80 and 200\,K on samples of the same batch \cite{Bertin24}.
A thorough analysis of the x-ray diffraction data supports the high sample quality with 
no indication of a significant amount of vacancies, see also Ref.\ \cite{Bertin24}. 
Measurements of the magnetic susceptibility $\chi$ and the specific heat $C_p$ 
(see Fig.\ \ref{fig:cp_chi}) yield results that agree with 
previous reports \cite{Cooke59,Khan19,Reig20,Bhaskaran21}, e.g., concerning the 
N\'eel temperature $T_N$\,=\,3.1\,K, 
the Weiss temperature $\theta_{\rm W}$\,=\,\mbox{-43.9\,K}, 
the sizable frustration parameter $f$\,=\,$|\theta_{\rm W}|/T_N$\,$\approx$\,14, 
and the effective magnetic moment $\mu_{\rm eff}$\,=\,1.69\,$\mu_{\rm B}$. The latter is 
in good agreement with the value 
$2\sqrt{j(j+1)}$\,$\mu_{\rm B}$\,$\approx$\,1.73\,$\mu_{\rm B}$ 
expected for an ideal $j$\,=\,1/2 ground state in a cubic environment.

RIXS spectra were measured at the Ir $L_3$ edge at beamline ID20 of the European Synchrotron 
Radiation Facility. In order to resonantly enhance the spin-orbit exciton, we 
tuned the energy of the incident photons to 11.214\,keV 
where an energy resolution of 25\,meV was achieved \cite{Moretti13,Moretti18}. 
RIXS data were collected at 10, 100, 200, and 300\,K on a (111) surface with (001) and (110) 
lying in the horizontal scattering plane. The incident x-ray photons were $\pi$ polarized. 
The RIXS spectra at different temperatures have been normalized by the spectral weight of 
the spin-orbit exciton. The transferred momentum $\mathbf{q}$ is given in reciprocal lattice 
units. 
Furthermore, we study the linear optical response of K$_2$IrCl$_6$ in the range from 0.1 
to 6\,eV, i.e., from the infrared up to the UV.\@ 
Using a Woollam VASE ellipsometer from 1 to 6\,eV, we address the optical response above 
the Mott gap at 300\,K.\@ For cubic symmetry, such ellipsometric measurements directly yield 
the optical conductivity $\sigma_1(\omega)$ \cite{Azzam87}. 
Due to inversion symmetry on the Ir site, the spin-orbit exciton in optics corresponds to 
a parity-forbidden excitation. However, it acquires finite spectral weight in a phonon-assisted process. 
For frequencies below the Mott gap, measurements of the infrared transmittance $T(\omega)$ 
are ideally suited to study such weak absorption features in transparent single crystals of 
appropriate thickness \cite{Rueckamp05}, as recently demonstrated on the sister compounds 
K$_2$OsCl$_6$ and K$_2$ReCl$_6$ \cite{Warzanowski23,Warzanowski24}. 
For K$_2$IrCl$_6$, we studied single-crystalline samples with a thickness $d$\,=\,380(8)\,\si{\micro\meter}, 95(3)\,\si{\micro\meter}, and 46(3)\,\si{\micro\meter}. 
Infrared transmittance data were measured from 0.15 to 1.85\,eV with an energy resolution 
of 8\,cm$^{-1}$ ($\approx$\,1\,meV) using a Bruker IFS\,66 v/S Fourier-transform infrared 
spectrometer equipped with a continuous-flow $^4$He cryostat.

\section{Spin-orbit exciton in RIXS}

\begin{figure}[t]
	\centering
	\includegraphics[width=\columnwidth]{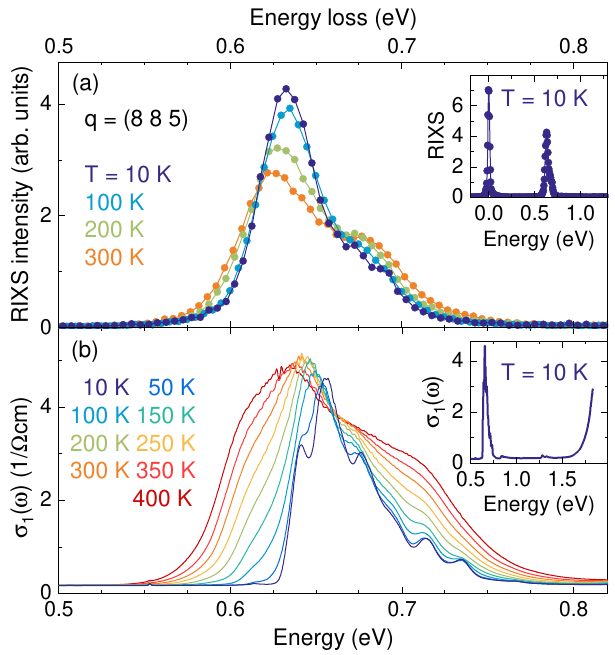}
	\caption{Spin-orbit exciton in $L_3$-edge RIXS and optics. 
	(a) RIXS spectra at $\mathbf{q}$\,=\,(8\,\,8\,\,5) show 
	a two-peak structure around 0.63\,eV.\@ 
	Inset: RIXS data over the entire measured range. The spin-orbit exciton is the only 
	inelastic feature.
	(b) Optical conductivity $\sigma_1(\omega)$ in the same energy range as used 
	in (a). The shift of the peak energy compared to RIXS at 10\,K, the existence of 
	several sidebands, and the increase of the spectral weight with increasing temperature 
	highlight the phonon-assisted character.
	Inset: The steep onset of excitations across the Mott gap is observed at about 1.7\,eV at 10\,K.}
	\label{fig:data_all}
\end{figure}

\begin{figure}[t]
	\centering
	\includegraphics[width=\columnwidth]{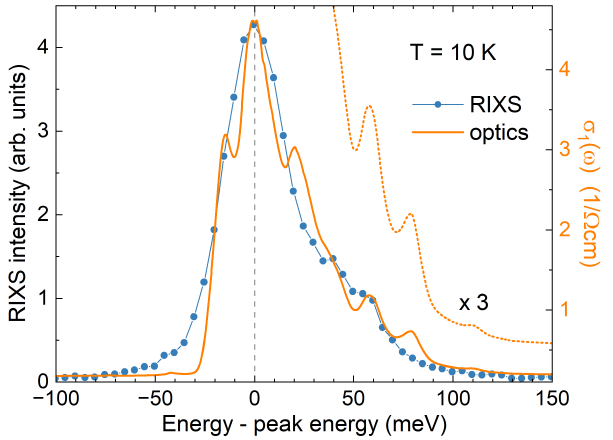}
	\caption{Spin-orbit exciton at 10\,K.\@  The overall line shape is similar in 
	RIXS (blue) and optics (orange, right axis). 
	The optical conductivity shows a superposition of phonon-assisted excitations 
	with different phonon energies, which effectively mimics the larger energy 
	resolution of 25\,meV in RIXS (see main text). 
	For the comparison, both data sets have been shifted by the 
	respective peak energy (633\,meV for RIXS, 656\,meV for $\sigma_1(\omega)$)
	to roughly compensate for the phonon shift.
    Dotted line: $\sigma_1(\omega)$ multiplied by 3 to emphasize the weak features 
    at 0.735 and 0.766\,eV, which are 0.08 and 0.11\,eV above the main peak.}
	\label{fig:RIXS_optics_shift}
\end{figure}

The hallmark excitation of a local $j$\,=\,1/2 ground state is the spin-orbit exciton, i.e., 
the excitation to the $j$\,=\,3/2 quartet that is expected at $1.5 \lambda$ in cubic symmetry 
\cite{Jackeli09,Moretti14CEF}, where $\lambda$ is the spin-orbit coupling constant, 
see sketch in Fig.\ \ref{fig:structure}(c).
RIXS data of K$_2$IrCl$_6$, measured up to 1.3\,eV with transferred momentum 
$\mathbf{q}$\,=\,(8\,\,8\,\,5), show the spin-orbit exciton around 0.63\,eV, see Fig.\ 
\ref{fig:data_all}(a). The absence of further inelastic features in this energy range 
demonstrates the pure Ir$^{4+}$ valence 
of the compound, see inset of Fig.\ \ref{fig:data_all}(a). 
The narrow line width is typical for local, weakly interacting $j$\,=\,1/2 moments 
\cite{Liu12,Gretarsson13,Rossi17,Aczel19,Revelli19,Ruiz21,delaTorre21,Jin22,Magnaterra23Ti,delaTorre23}, 
while broader, dispersive features are observed in compounds with larger exchange interactions 
such as Sr$_2$IrO$_4$ and Sr$_3$Ir$_2$O$_7$ \cite{Kim12,Kim14,Lu18,Kim12327,Moretti15}.
In agreement with previous RIXS data on K$_2$IrCl$_6$ measured with 35\,meV resolution 
\cite{Reig20}, we observe a two-peak structure of the spin-orbit exciton. 
The main peak at 0.63\,eV features a weak shoulder that is about 0.05\,eV 
higher in energy, see Fig.\ \ref{fig:RIXS_optics_shift}. 
The relative intensity of this shoulder increases with increasing temperature, see Fig.\ \ref{fig:data_all}(a), 
in agreement with previous results for 10 and 300\,K \cite{Reig20}. 
Similar two-peak structures were reported in $L$-edge RIXS on the sister compounds K$_2$IrBr$_6$ 
and (NH$_4$)$_2$IrCl$_6$, also in their cubic phases \cite{Reig20,Khan21}. 
Furthermore, a splitting of the spin-orbit exciton in K$_2$IrCl$_6$ has been observed in 
Raman scattering \cite{Lee22} and in infrared absorption at room temperature \cite{Meggle23}.

In fact, such a splitting of the spin-orbit exciton is a common feature in RIXS studies 
on Ir$^{4+}$ compounds, see, e.g., Refs.\ \cite{Liu12,Gretarsson13,Rossi17,Aczel19,Revelli19,Ruiz21,delaTorre21,Jin22,Magnaterra23Ti,delaTorre23,Kim12,Kim14,Lu18,Kim12327,Moretti15}.
Typically, the splitting is attributed to a noncubic crystal field contribution $\Delta_{\mathrm{CF}}$ 
that lifts the degeneracy of the $j$\,=\,3/2 quartet.
For a single site and assuming a tetragonal distortion, the physics is described by \cite{Moretti14CEF}
\begin{align}
\mathcal{H}_{\mathrm{single}}=\lambda \mathbf{S}\cdot\mathbf{L}+\Delta_{\mathrm{CF}}L^2_z
\label{eq:single-site}
\end{align}
where $L_z$ is the component of the angular moment \textbf{L} along the tetragonal axis. 
A finite $\Delta_{\mathrm{CF}} \ll \lambda$ lifts the degeneracy of the 
$j$\,=\,3/2 quartet, resulting in an experimental peak splitting $\Delta_{\rm exp}$\,=\,$\frac{2}{3}\Delta_{\rm CF}$. 
Following this scenario, we empirically fit the RIXS spectra with a sum of two Voigt profiles. 
At 10\,K, the fit yields peak energies of 0.635 and 0.676\,eV, see Appendix B.\@
Solving Eq.\ \eqref{eq:single-site}, we find $\lambda$\,=\,434(1)\,meV and two possible values  $\Delta_{\mathrm{CF}}$\,=\,62\,meV and \mbox{-58\,meV} for elongation and compression of the 
octahedra, respectively.

However, a finite value of the noncubic crystal field splitting $\Delta_{\rm CF} \neq 0$ is 
in conflict with the globally cubic structure observed in x-ray and neutron diffraction 
experiments \cite{Khan19,Reig20,Bertin24} 
as well as with the isotropic $g$ factor found in an ESR  study \cite{Bhaskaran21}. 
There are two scenarios to resolve this apparent discrepancy.  
(i) Cubic symmetry is broken locally in the initial state of the excitation process. 
(ii) A vibronic character of the Jahn-Teller active $j$\,=\,3/2 excited states 
\cite{Iwahara23,Plotnikova16,Streltsov20} yields sidebands while the cubic symmetry 
of the ground state is preserved.

\begin{figure}[t]
	\centering
	\includegraphics[width=\columnwidth]{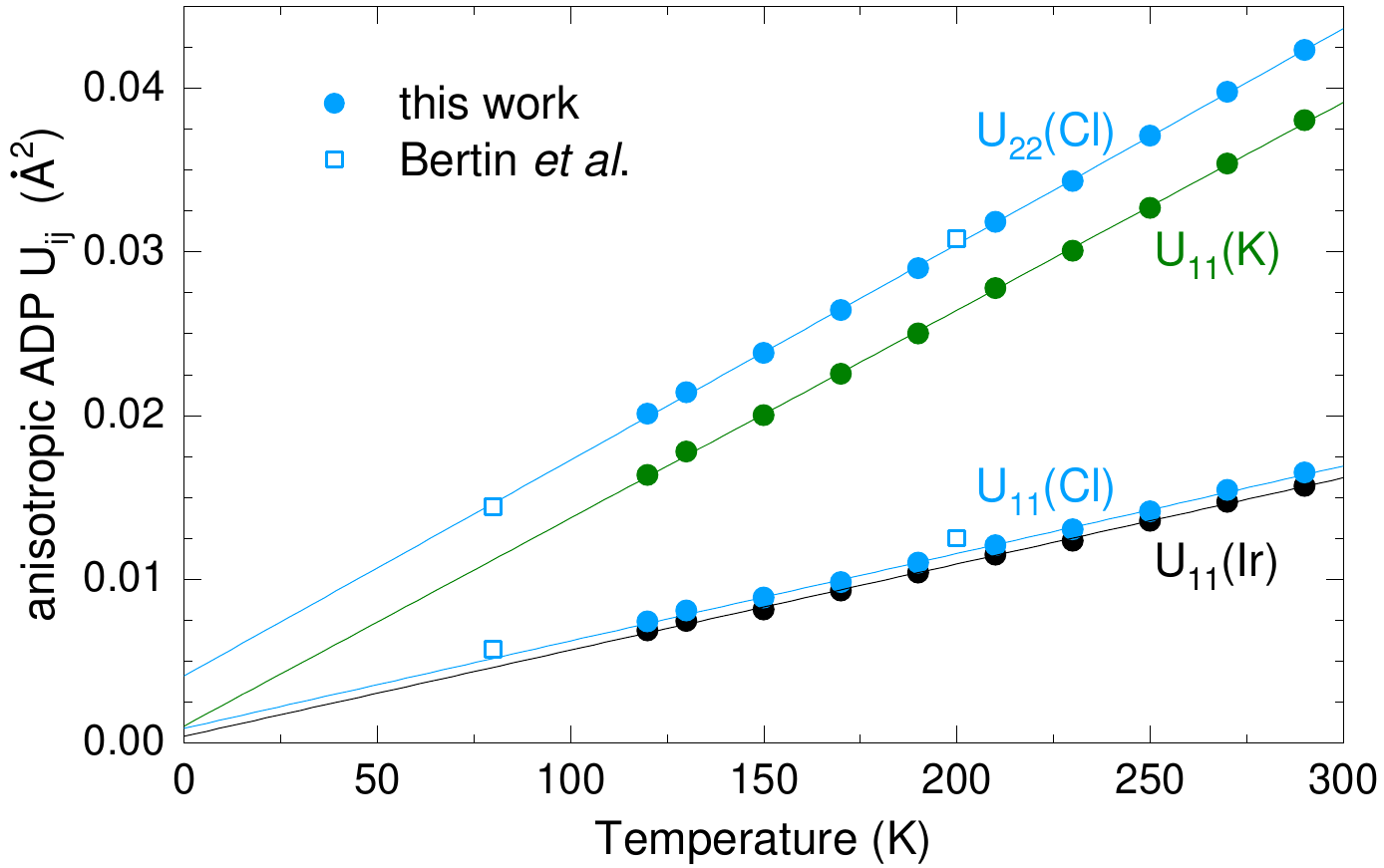}
	\caption{Anisotropic atomic displacement parameters (ADP) $U_{ij}$ as determined 
	by single-crystal x-ray diffraction. 
	Error bars are plotted but are smaller than the symbol size. 
	Data at 80 and 200\,K (open symbols) have been taken from Bertin \textit{et al.}\ \cite{Bertin24}. 
	The lines are guides to the eye. }
	\label{fig:ADP}
\end{figure}

In scenario (i), local deviations from cubic symmetry may be caused by either 
static defects or a strong magneto-elastic coupling that triggers distortions \cite{Revelli19}. 
The latter can be reconciled with global cubic symmetry if the correlation length is small. 
Based on the short time scale of the RIXS process, also the thermal occupation of 
low-lying phonon modes effectively may break the cubic symmetry on the Ir site in 
the initial state \cite{Khan21}. 
In fact, x-ray studies reported comparably large atomic 
displacement parameters (ADPs) in the K$_2$PtCl$_6$-type antifluorite halides $A_2MX_6$ 
in general and also in K$_2$IrCl$_6$ \cite{Khan19,Bertin24}. 
Using a cubic crystal structure in the analysis of the x-ray data, such large ADPs 
in general may reflect local disorder or dynamical effects such as the thermal population 
of low-energy phonon modes. 
Based on a thorough analysis of the x-ray diffraction data, Bertin \textit{et al.}\ 
\cite{Bertin24} conclude that there is no indication for local disorder in K$_2$IrCl$_6$. 
Note that neighboring $MX_6$ or IrCl$_6$ octahedra are not connected in $A_2MX_6$ or 
K$_2$IrCl$_6$, see Fig.\ \ref{fig:structure}(a), i.e., they do not share a common ligand.  
This causes a rotational instability and low-energy librational modes
\cite{Khan19,Lynn78,Mintz79,Oleary70}. 
For the halide ions, the largest ADPs indeed are found perpendicular to the 
Ir-Cl bonds \cite{Bertin24,Khan19}, as expected for a predominantly rigid rotation of the 
octahedra in contrast to a local, noncubic distortion away from octahedral symmetry. 
To scrutinize this scenario of a dynamical origin for our samples, we studied the 
temperature dependence of the ADPs $U_{ij}$ of the Ir, K, and Cl ions in the range from 
120 to 290\,K via single-crystal x-ray diffraction measurements, see Appendix A.\@
The results are plotted in Fig.\ \ref{fig:ADP} and agree very well with previous data 
reported for 80 and 200\,K that were collected on samples from the same batch \cite{Bertin24}. 
For the Cl ions, $U_{11}$ and $U_{22}$ refer to the displacement along and 
perpendicular to an Ir-Cl bond, respectively. 
We find $U_{22}{\rm (Cl)} \gg U_{11}{\rm (Cl)}$, in agreement with 
Refs.\ \cite{Khan19,Bertin24}. 
Using $\tan \varphi$\,=\,$\sqrt{U_{22}({\rm Cl})}/d$, we estimate the largest rotation 
angle $\varphi$\,$\approx$\,5$^\circ$ at 300\,K.\@ 
The temperature dependence strongly supports a predominantly dynamical origin of the 
large $U_{ij}$. Above 120\,K, we find a linear behavior for all $U_{ij}$ that extrapolates 
to small values at low temperature, in agreement with data reported by 
Khan \textit{et al.}\ \cite{Khan19}. The data do not provide any evidence for 
a static contribution of local distortions or defects. 
A more precise quantitative determination of a possible small static contribution present 
already at very low temperatures requires to consider zero point fluctuations 
\cite{Schweiss94} and a possible temperature dependence of the phonon energies. 
This is beyond the scope of our study. 
In comparison, the cubic double perovskite Ba$_2$CeIrO$_6$ exhibits a much smaller 
temperature dependence of the ADPs, with $U_{22}$(O)\,=\,0.022\,\AA$^2$ at 100\,K 
and 0.024\,\AA$^2$ at 300\,K \cite{Revelli19}. This suggests static disorder 
that may be explained by the presence of a few percent of Ce-Ir site disorder \cite{Revelli19}. 
In K$_2$IrCl$_6$, the pronounced temperature dependence of the $U_{ij}$ is in striking contrast 
to the much smaller temperature dependence of the peak splitting in RIXS.\@ 
We conclude that the sizable atomic displacement parameters at elevated temperature 
have a dynamical origin. 
In particular, $U_{22}$(Cl) predominantly can be attributed to rigid rotations of the 
octahedra.

Thermal population of low-energy rotations of the octahedra preserves cubic symmetry 
on average. For spectroscopy, however, it may break the cubic symmetry on the Ir site 
if the time scale of the electronic excitation is short enough. 
Nevertheless, only a small effect is expected for a rigid rotation of the octahedra. 
More precisely, Khan \textit{et al.}\ \cite{Khan21} predict 
a dynamical rotation angle of about 5$^\circ$ in the cubic phase of the sister compound 
K$_2$IrBr$_6$ above 170\,K, similar to the value that we find for K$_2$IrCl$_6$, 
as discussed above. This, however, is expected to cause a noncubic crystal-field 
splitting of less than 10\,meV at high temperature \cite{Khan21}, which is by far 
not large enough to explain the splitting observed in RIXS.\@ In particular, a thermal 
population of low-energy octahedral rotations cannot explain the presence of high-energy 
sidebands already at 10\,K.

A survey of many compounds of the $AMX_6$ antifluorite family reveals a clear correlation 
between the Goldschmidt tolerance factor $t$ and the structural transition temperature $T_s$ 
from the high-temperature cubic phase to a phase with lower symmetry \cite{Bertin24}. 
The Goldschmidt tolerance factor $t$ is based on the atomic radii and usually 
provides a criterion for rotational phase transition in perovskites. 
For the antifluorite halides, Bertin \textit{et al.}\ \cite{Bertin24} find a critical 
value of $t \approx 1$ below which a rotational phase transition occurs, where $T_s$ 
increases linearly with $t$ decreasing below 1. For K$_2$IrCl$_6$, the tolerance factor 
amounts to 1.0019 \cite{Bertin24}, in agreement with the observation of 
cubic symmetry down to at least 0.3\,K \cite{Khan19,Reig20}.

Local distortions may also result from strong magneto-elastic coupling that has been proposed 
based on very similar results for the double perovskite Ba$_2$CeIrO$_6$ \cite{Revelli19}, 
another compound with $j$\,=\,1/2 moments forming an fcc lattice. 
Also Ba$_2$CeIrO$_6$ shows global cubic symmetry in x-ray diffraction measurements 
while RIXS reveals a splitting of the spin-orbit exciton of about 0.1\,eV \cite{Revelli19,Aczel19}, 
roughly a factor two larger than in K$_2$IrCl$_6$. 
Exchange couplings on the fcc lattice are highly frustrated, and this frustration may be 
lifted by small distortions \cite{Revelli19}. Such distortions may escape detection 
in diffraction experiments if they are either very small or show a short correlation 
length. According to Iwahara and Furukawa \cite{Iwahara23}, 
an interpretation of the two-peak line shape of the spin-orbit exciton of K$_2$IrCl$_6$ 
in terms of a static crystal-field splitting requires a displacement of the Cl ligands of 
$\delta r$\,=\,0.007\,\AA. 
Such distortions would cause deviations from the cubic structure, but corresponding Bragg peaks 
have not been found \cite{Khan19,Reig20,Bertin24}. 
Using a cubic structure in the analysis of diffraction data, deviations from cubic symmetry 
can also be detected via anomalously large atomic displacement parameters. 
However, effects as small as $(\delta r)^2 < 10^{-4}$\,\AA$^2$ cannot be detected in the ADPs. 
Altogether, the structural data do not provide any evidence for deviations from cubic symmetry. 
On the contrary, a thorough analysis of the x-ray diffraction data (see also Ref.\  \cite{Bertin24}) 
and the temperature dependence of the ADPs rather support cubic symmetry. 
Small distortions nevertheless cannot be excluded.

Scenario (ii) explains the two-peak structure of the RIXS spectra in terms of a 
vibronic character of the spin-orbit exciton, as recently proposed for K$_2$IrCl$_6$ \cite{Iwahara23}. 
Vibronic excitations emerge from the coupling between electronic and vibrational excitations 
\cite{Henderson89,Figgis,Ballhausen}.
An electronic excitation such as the spin-orbit exciton may change the charge-density 
distribution such that the lattice is not in its corresponding ground state, causing a 
series of phonon sidebands of a given electronic excitation. 
A basic example is given by the phonon satellite lines observed 
in photoemission on H$_2$ molecules \cite{Sawatzky89}. Suddenly removing an electron changes 
the equilibrium distance and leaves the molecule in a vibrationally excited state, i.e., 
the molecule rings. In K$_2$IrCl$_6$, this vibronic scenario does not break cubic symmetry 
in the $j$\,=\,1/2  ground state.

With the simple two-peak structure of the RIXS data, an unambiguous distinction between 
the two scenarios is challenging. 
To resolve the origin of the splitting in K$_2$IrCl$_6$, we study the spin-orbit exciton 
with optical spectroscopy, making use of the excellent energy resolution.
We will show that a vibronic character is the key to understand the peculiar spectra 
of K$_2$IrCl$_6$ in RIXS and optics. 
The vibronic sidebands reflect the hybridization of spin-orbit exciton and 
vibrational states, i.e., spin-orbital-lattice entanglement.

\section{Spin-orbit exciton in optical data}

\subsection{Phonon-assisted character}

The optical conductivity $\sigma_1(\omega)$ in the energy range of the spin-orbit exciton 
is plotted in Fig.\ \ref{fig:data_all}(b). The spin-orbit exciton is an intra-$t_{2g}$ 
excitation. 
In $\sigma_1(\omega)$, such on-site $d$-$d$ excitations are parity forbidden due to 
the presence of inversion symmetry on the Ir site. 
Finite spectral weight arises in a phonon-assisted process 
via the simultaneous excitation of an odd-parity phonon \cite{Figgis,Henderson89,Ballhausen,Rueckamp05}, 
which explains the complex line shape. 
Such weakly dipole-active excitations can be studied very well in the transparency window above 
the phonons and below the Mott gap, as reported recently for K$_2$ReCl$_6$ and K$_2$OsCl$_6$ 
\cite{Warzanowski23,Warzanowski24} 
as well as for the spin-orbit exciton in $\alpha$-RuCl$_3$ \cite{Warzanowski20}. 
In K$_2$IrCl$_6$ at 10\,K, we observe the onset of excitations across the Mott gap 
around 1.7\,eV, see inset of Fig.\ \ref{fig:data_all}(b). 
At the peak of the spin-orbit exciton, $\sigma_1(\omega)$ reaches values of about 5\,($\Omega$cm)$^{-1}$, 
which is roughly two orders of magnitude smaller than for the directly dipole-allowed intersite excitations 
$|d_i^5 d_j^5\rangle \rightarrow |d_i^4 d_j^6\rangle$ across the Mott gap, see Appendix C.\@

At 10\,K, the phonon-assisted character is evident from the shift of the peak energy in $\sigma_1(\omega)$ 
compared to RIXS and from the increase of the spectral weight with increasing 
temperature \cite{Hitchman79,Warzanowski24}. 
Despite the different excitation mechanisms, the overall line shape 
of the spin-orbit exciton is very similar in RIXS and $\sigma_1(\omega)$ at 10\,K.\@
This is highlighted in Fig.\ \ref{fig:RIXS_optics_shift}, 
where both data sets have been shifted by the peak energy to roughly compensate for the phonon shift. 
The RIXS data have been measured with an energy resolution of $\delta E$\,=\,25\,meV and show a 
strong main peak and smaller intensity at higher energy. 
Qualitatively, the optical data show a very similar behavior but resolve an additional fine structure. 
The corresponding subbands are due to a superposition of phonon-assisted processes for different 
symmetry-breaking phonon modes with phonon energies in the range of roughly 10 to 40\,meV, 
as discussed below. 
This superposition enhances the overall line width despite the much better energy resolution. 
The assignment of the subbands provides the key to understand the character of the weaker high-energy features. 
To this end, we need to have a close look at the peak assignment in the optical data.

\subsection{Assignment of phonon-assisted excitations}

\begin{figure}[t]
	\centering
	\includegraphics[width=\columnwidth]{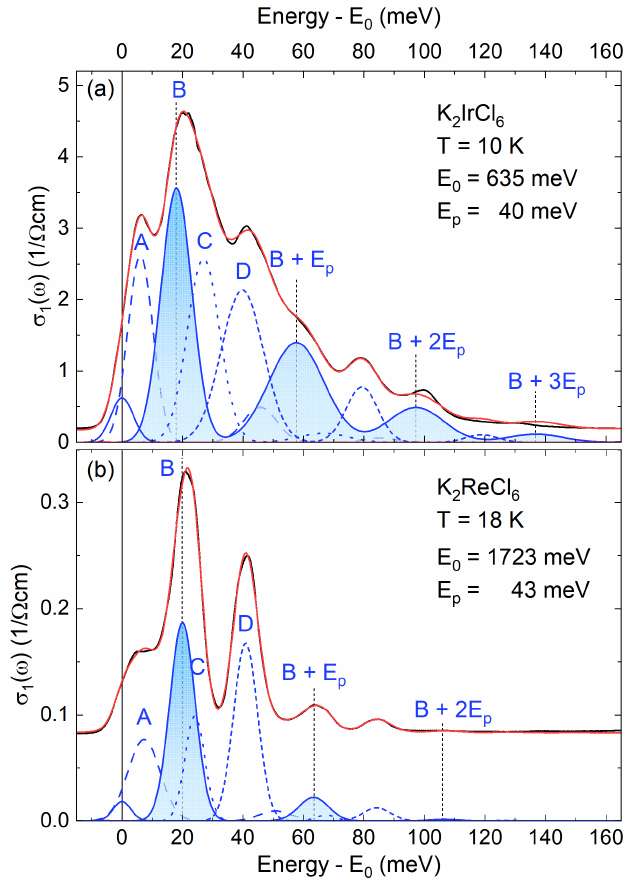}
	\caption{Comparison of the line shape of $\sigma_1(\omega)$ for phonon-assisted intra-$t_{2g}$ 
	excitations in (a) K$_2$IrCl$_6$ and (b) its $t^3_{2g}$ sister compound K$_2$ReCl$_6$ \cite{Warzanowski24}.  
	The data (black) are shifted by the respective bare electronic energy $E_0$. 
	The similar line shape underlines the common motif of four phonon-assisted excitations 
	A--D, each dressed by a vibronic sideband progression (see main text). For peak B, 
	this motif is emphasized by the blue shading. 
	Note that A--D refer to processes involving odd-symmetry phonons that yield a finite 
	spectral weight while $E_{\rm p}$ 
	is the energy of an even mode, see Eq.\ \eqref{eq:vib} and Fig.\ \ref{fig:FC}. 
	The weak mode at $E_0$ can be attributed to a magnetic dipole transition. 
	The red line depicts the sum of the blue curves plus a small constant offset. 
	}
	\label{fig:shift}
\end{figure}

At 10\,K, the dominant phonon-assisted peaks in $\sigma_1(\omega)$ occur at $E_0 + E_{\mathrm{odd}}$, 
where $E_0$\,=\,635\,meV is the energy of the bare electronic spin-orbit exciton while $E_{\mathrm{odd}}$ 
denotes the energy of a symmetry-breaking phonon mode. 
The cubic crystal structure of K$_2$IrCl$_6$ hosts four odd-symmetry optical phonon modes 
that have been observed at 10, 18, 23, and 41\,meV in infrared spectroscopy and/or 
inelastic neutron scattering \cite{Adams63,Bottger72,Meggle23,Parker98}. 
Given the quasimolecular crystal structure with separate IrCl$_6$ octahedra, the latter three of 
these modes can be viewed as the three odd-symmetry normal modes of a single IrCl$_6$ octahedron 
\cite{Henderson89} while the fourth one is a lattice phonon mode. 
The main peaks in $\sigma_1(\omega)$ are well described by considering 
$E_{\mathrm{odd}}$\,=\,6, 18, 27, and 40\,meV, 
see peaks A--D in Fig.\ \ref{fig:shift}(a). 
Taking into account that the symmetry-breaking modes do not have to be at the $\Gamma$ point, 
the energies show good agreement with the values reported above, lending strong support to our peak assignment.

With increasing temperature, the spectral weight of a phonon-assisted process 
at $E_0 + E_{\rm odd}$ increases with $1+n(T)$, where $n(T)$ denotes the phonon 
occupation number of the odd-symmetry mode \cite{Henderson89}. 
At finite temperatures, also phonon annihilation processes occurring at 
$E_0 - E_{\rm odd}$ acquire spectral weight in $\sigma_1(\omega)$ that increases 
proportional to $n(T)$. 
The phonon-assisted character of the spin-orbit exciton in $\sigma_1(\omega)$ 
hence explains the different temperature dependence observed in RIXS and optics, 
in particular the overall increase of the spectral weight in $\sigma_1(\omega)$ 
with increasing temperature as well as the pronounced enhancement of absorption below $E_0$.

In the chloride, the energy $E_{\rm odd}$ of the symmetry-breaking phonons is limited 
to about 40\,meV.\@ The phonon-assisted character in $\sigma_1(\omega)$ hence explains 
the dominant spectral weight and the subbands in the range of roughly 0.64 to 0.68\,eV.\@ 
This does not yet reveal the nature of the weak features at higher energies, e.g., at 0.71, 
0.74, and 0.77\,eV, reaching energies as high as about $E_0 + 0.1$\,eV.\@ 
In general, a phonon-assisted character of the spin-orbit exciton is expected in 
$\sigma_1(\omega)$ as long as there is inversion symmetry on the Ir site. 
This phonon-assisted picture successfully has been used to describe the equivalent peaks 
of IrCl$_6$ impurities in different $AMX_6$-type host compounds such as cubic Cs$_2$ZrCl$_6$ 
or K$_2$SnCl$_6$ \cite{Keiderling75,Yoo86}. 
The data of these IrCl$_6$ impurities reveal the existence of vibronic sidebands. 
Remarkably, both the observed energies and the overall structure are very similar 
to our observations in single crystalline K$_2$IrCl$_6$.

\section{Vibronic excitations}

\subsection{Vibronic sidebands and the Franck-Condon principle}

A vibronic character emerges from the coupling of electronic excitations  to vibrational 
degrees of freedom. An on-site intra-$t_{2g}$ excitation may give rise to a change of 
the charge distribution such that the lattice is not in its ground state anymore. 
This yields a series of phonon sidebands of the electronic excitation 
\cite{Henderson89,Figgis,Ballhausen} 
and has been claimed to describe the RIXS data of K$_2$IrCl$_6$ \cite{Iwahara23}.

\begin{figure}[t]
	\centering
	\includegraphics[width=\columnwidth]{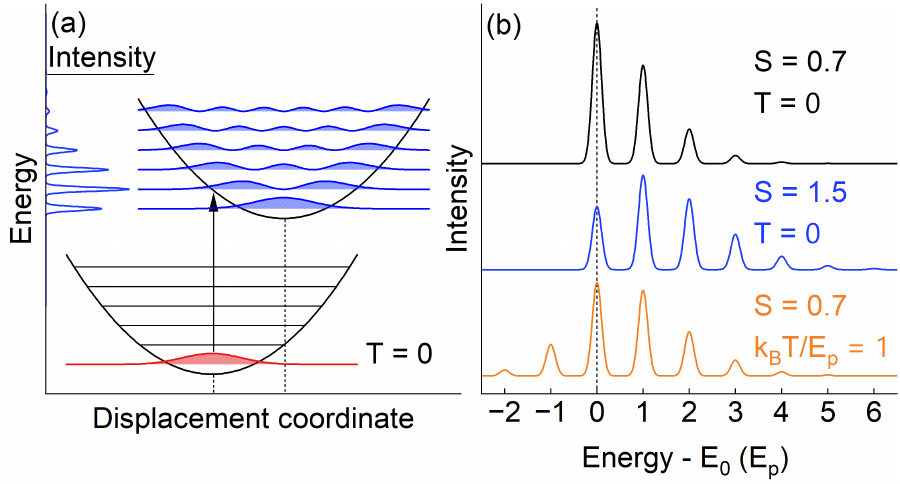}
	\caption{Sketch of vibronic excitations in optical absorption according to the 
	Franck-Condon principle.
	(a) Excitation from the ground state at $T$\,=\,0
	(red) to vibronic excited states (blue). 	
	The two parabolas depict the harmonic lattice potential with phonon excitations, 
	where the shaded area gives the squared amplitude of the wavefunction. 
	The parabola of the electronic excited state is shifted horizontally since a change 
	of the charge distribution affects the equilibrium positions of the ions 
	for finite electron-phonon coupling $g$. 
	The arrow corresponds to a "vertical" transition according to the Franck-Condon principle, 
	assuming a negligible timescale for electronic excitations. 
	On the left (blue lines), the evenly spaced comb of peaks with splitting $E_{\rm p}$ 
	illustrates the resulting excitation spectrum. 
	(b) Examples of the Franck-Condon line shape for different values of the Huang-Rhys 
	factor $S$, see Eq.\ \eqref{eq:vib}. 
	The orange line corresponds to a finite temperature $T$\,=\,$E_{\rm p}/k_{\rm B}$. 
	For clarity, the curves are offset vertically.
	}
	\label{fig:FC}
\end{figure}

In fact, vibronic sidebands of on-site intra-$t_{2g}$ excitations are a common feature 
in the optical conductivity of the $AMX_6$ family, e.g., in the sister compounds 
K$_2$ReCl$_6$ and K$_2$OsCl$_6$ \cite{Pross74,Yoo87,Bettinelli88,Kozikowski83,Warzanowski23,Warzanowski24}. 
In RIXS measurements on the Os and Re halides, however, the energy resolution was not sufficient 
to resolve the vibronic character \cite{Warzanowski23,Warzanowski24}. 
To put things into perspective, we consider the $5d^3$ compound K$_2$ReCl$_6$ that shows five 
different intra-$t_{2g}$ excitations \cite{Warzanowski24}. 
Optical data for one of them are depicted in Fig.\ \ref{fig:shift}(b). 
In $\sigma_1(\omega)$ of K$_2$ReCl$_6$, the subbands are very sharp and well resolved. 
The stronger peaks A--D reveal the subbands of the four odd phonon modes at 
$E_0+E_{\rm odd}$, demonstrating the phonon-assisted character \cite{Warzanowski24}. On top, 
weak vibronic sidebands are resolved at $E_0+E_{\rm odd}+ m E_{\rm p}$, where $m E_{\rm p}$ 
with integer $m$ denotes the energies of a phonon progression according to the 
Franck-Condon principle, see Fig.\ \ref{fig:FC}.
For the data of K$_2$ReCl$_6$ in Fig.\ \ref{fig:shift}(b), the electronic excited state is 
a $\Gamma_7$ doublet \cite{Warzanowski24}. An interpretation of the high-energy sidebands 
in terms of crystal-field splitting is hence not applicable. These features unambiguously 
are of vibronic character. Remarkably, a very similar sideband structure has been observed 
for all five intra-$t_{2g}$ excitations in K$_2$ReCl$_6$ \cite{Warzanowski24}. 
This common motif also applies to the data of K$_2$IrCl$_6$ in Fig.\ \ref{fig:shift}(a).

The Franck-Condon principle offers a simplified, analytic, and intuitive description 
of the vibronic line shape. It assumes that the timescale of electronic excitations is 
negligible compared to phonon timescales. 
For an optical absorption process, this corresponds to an instantaneous, "vertical" transition 
from the electronic ground state to the excited state, see Fig.\ \ref{fig:FC}(a), 
where "vertical" implies no change on the horizontal axis that denotes 
a generalized displacement coordinate. 
For illustration and simplicity, we assume that there is only one odd-symmetry mode 
with energy $E_{\rm odd}$ that is relevant for the optical conductivity but not for RIXS.\@ 
Furthermore, we assume that the electronic excited state is a Kramers doublet and 
that there is one dominant phonon mode with energy $E_{\rm p}$ 
that governs the progression of vibronic sidebands. 
Neglecting the entanglement between electronic and lattice degrees of freedom, 
the vibronic sidebands are shifted by $m E_{\rm p}$ in the Franck-Condon approximation, 
i.e., they yield an evenly spaced comb of peaks at 
$E_m$\,=\,$E_0 + E_{\rm odd} + m E_{\rm p}$ in $\sigma_1(\omega)$. 
At $T$\,=\,0, the line shape for optical absorption is described by \cite{Henderson89,Huang50}
\begin{eqnarray}
	\label{eq:vib0}
	I(E) \! & \! = \! & I_0 
	\,   \sum_{m = 0}^{\infty}\frac{e^{-S}\, S^m}{m!}\, \delta(E_0+E_{\rm odd}+mE_{\rm p}-E)
\end{eqnarray}
with $\sigma_1(E)$\,=\,$I(E)\cdot E$, where $I_0$ is proportional to the squared dipole matrix element 
of the (phonon-assisted) electronic transition, $e^{-S}\, S^m/m!$ is the Franck-Condon factor, 
and $S$ is the Huang-Rhys factor, which is equivalent to the average number of phonons that are present 
in the excited state. It is a measure of the electron-phonon coupling constant $g$ and governs 
the line shape. 
A large value of $S$ creates an envelope with a more Gaussian-like intensity distribution 
while a small $S$ yields an asymmetric envelope, see Fig.\ \ref{fig:FC}(b). 
At finite temperatures, one has to consider the thermal occupation of the sideband phonon modes, 
$n_p$\,=\,$1/\left\lbrack\exp(E_{\rm p}/k_{\mathrm{B}}T)-1\right\rbrack$. The sum hence has to 
include negative values of $m$ \cite{Henderson89,Huang50},
\begin{eqnarray}
	\nonumber
	I(E) & \! = \! & I_0 
	\sum_{m = -\infty}^{\infty}\left(\frac{n_p+1}{n_p}\right)^{m/2} J_{m}\!\left(2S\sqrt{ n_p(n_p\!+\!1)}\right)\\
	& \! \times \! & e^{-S(2n_p+1)} \,\, \, \delta(E_0+E_{\rm odd}+mE_{\rm p}-E),
	\label{eq:vib}
\end{eqnarray}
where $J_m$ is the modified Bessel function of $m^{\rm th}$ order. 
For comparison with experiment, we replace the $\delta$ function in Eq.\ \eqref{eq:vib} 
with a Gaussian profile. 
This yields excellent agreement with $\sigma_1(\omega)$ of $5d^3$ K$_2$ReCl$_6$, see Fig.\ \ref{fig:shift}(b). 
The experimental spectrum can be described by a superposition of 
such Franck-Condon progressions for all four odd-symmetry phonon modes. 
In K$_2$ReCl$_6$, most of the spectral weight is in the peaks of order $m$\,=\,0 with only a 
small contribution of $m$\,=\,1 and basically negligible spectral weight for $m$\,=\,2. 
This intensity distribution corresponds to a small Huang-Rhys factor $S$\,=\,0.14. 
This reflects the spin-forbidden character of the excitation in K$_2$ReCl$_6$ that is observed 
at about $5J_H$, where $J_H$ denotes Hund's coupling. The excitation roughly can be viewed as a 
flip of the spin of one electron in the $t_{2g}^3$ configuration, which causes only a small change 
of the charge distribution and corresponds to a small $S$. 
The same approach also yields an excellent description of $\sigma_1(\omega)$ of 
K$_2$IrCl$_6$, see Fig.\ \ref{fig:shift}(a). The blue shading highlights the 
Franck-Condon sidebands of peak B with a larger but still small value of $S$\,$\approx$\,0.7. 
The larger $S$ yields detectable spectral weight of the peaks of order $m$\,=\,2 and even 3. 
Thereby, the vibronic model naturally explains the large number of peaks and the existence of 
weak features at energies as high as 0.74 and 0.77\,eV, 
more than 0.1\,eV above the bare electronic excitation energy $E_0$\,=\,635\,meV.\@

\begin{figure}[t]
	\centering
	\includegraphics[width=\columnwidth]{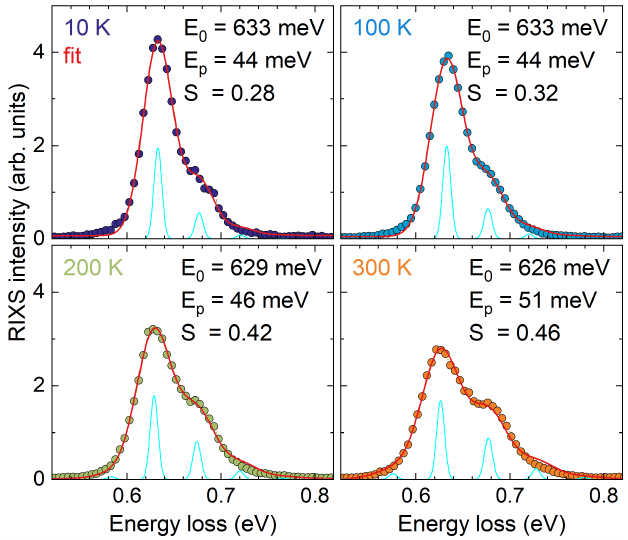}
	\caption{Vibronic fits of the RIXS data at different temperatures using the 
	Franck-Condon picture from Eq.\ \eqref{eq:vib}. In each panel, we state 
	the energy $E_0$ of the bare electronic excitation, 
	the energy $E_{\mathrm{p}}$ of the phonon progression mode, 
	and the Huang-Rhys factor $S$. 
    The red lines depict the full fit. 
    Cyan curves illustrate the phonon progression of sidebands for the same parameters 
    with 10\,meV width of the Gaussian profile and reduced total intensity. 
}
	\label{fig:rixs_fit}
\end{figure}

Thus far we discussed the Franck-Condon approximation for an optical excitation. 
We claim that Eq.\ \eqref{eq:vib} can also be applied to vibronic sidebands 
of an electronic excitation studied in direct RIXS, which is supported by the overall 
agreement of the line shapes shown in Fig.\ \ref{fig:RIXS_optics_shift}. 
Note that the case is different for elementary phonons and multi-phonons. These are 
excited in an \textit{indirect} RIXS process that can be approximated by using 
two Franck-Condon factors \cite{Ament11EPL,Gilmore23,Geondzhian20,Braicovich20}, 
see Sec.\ \ref{sec:phonon}.
For K$_2$IrCl$_6$, Eq.\ \eqref{eq:vib} provides an excellent empirical description of our 
RIXS spectra, see Fig.\ \ref{fig:rixs_fit}. 
At 10\,K, the fit yields $S$\,=\,0.30, $E_{\rm p}$\,=\,44\,meV, 
and the bare electronic energy $E_0$\,=\,633\,meV.\@ 
With an energy resolution of 25\,meV, the latter agrees very well with $E_0$\,=\,635\,meV 
found in $\sigma_1(\omega)$. In the next paragraph, we discuss $E_p$ to collect evidence 
for the spin-orbital-lattice entangled character.

\subsection{Beyond the Franck-Condon approximation}

The Franck-Condon principle is valuable for an intuitive explanation of the overall line shape. 
For a microscopic description of K$_2$IrCl$_6$, Iwahara and Furukawa \cite{Iwahara23} 
treated electron-phonon coupling $g$ and spin-orbit coupling $\lambda$ on the same footing. 
They find that the spin-orbit exciton with its vibronic sidebands is of spin-orbital-lattice 
entangled character. Strictly speaking, the simple picture of the simultaneous excitation of 
a spin-orbit exciton and one or several phonons, causing a comb of equidistant peaks, 
is only applicable for $g\rightarrow 0$, in which case the intensity of the sidebands vanishes. 
For finite $g$, the excitation spectrum is more complex. Iwahara and Furukawa \cite{Iwahara23}
predict a larger number of peaks but many of them are small and hard to resolve experimentally. 
Furthermore, hybridization shifts the energies such that the sidebands are not equidistant 
anymore. In K$_2$IrCl$_6$, however, we can only determine two peak energies in RIXS, 
and the complex line shape of the phonon-assisted excitations in $\sigma_1(\omega)$ 
does not allow us to clearly identify deviations from a comb-like behavior. 
The hybridized character may also give rise to differences in line shape between RIXS and $\sigma_1(\omega)$, 
e.g., due to the presence of the symmetry-breaking phonon mode in the case of the optical data. 
However, we are lacking a theoretical prediction for such differences. 
From an experimental point of view, the hybridized character at this stage can be detected 
in the value of the peak splitting, which is captured by $E_{\rm p}$ in Eq.\ \eqref{eq:vib}.

At 10\,K, the fits yield $E_{\rm p}$\,=\,40\,meV in $\sigma_1(\omega)$ and 44\,meV 
in RIXS, see Figs.\ \ref{fig:shift}(a) and \ref{fig:rixs_fit}. 
In general, the even electronic excitations may couple to phonons of $a_{1g}$, $e_g$, 
and $t_{2g}$ symmetry \cite{Black75}, 
where the latter usually is neglected. The elementary phonon modes of $a_{1g}$ and $e_g$ 
symmetry have been observed in Raman scattering at 44\,meV and 37\,meV, 
respectively \cite{Lee22}. In the Franck-Condon approximation, the experimental splitting 
thus appears to suggest a predominant coupling to the $a_{1g}$ mode, the breathing mode of 
the IrCl$_6$ octahedron. This applies if, e.g., both the electronic ground state 
and the excited state show cubic symmetry, which is the case for ideal $j$\,=\,1/2 and 3/2 
states. For the doublet excited state of K$_2$ReCl$_6$ addressed in Fig.\ \ref{fig:shift}(b), 
the phonon sidebands with $E_{\rm p}$\,=\,43\,meV have been attributed to the $a_{1g}$ mode \cite{Yoo87,Warzanowski24}. 
However, the situation is different in K$_2$IrCl$_6$ due to 
the degeneracy of the excited $j$\,=\,3/2 quartet that is lifted by coupling to the 
Jahn-Teller active $e_g$ phonon mode \cite{Iwahara23}. 
In this case, the energy splitting between the lowest two strong RIXS peaks 
roughly is given by $E_{\rm p}\cdot (1+g^2/8)$. 
Note that $g$ is dimensionless in the definition of Ref.\ \cite{Iwahara23}. 
With $E_{\rm p}$\,=\,37\,meV for the $e_g$ mode and $g$\,=\,1.2, 
Iwahara and Furukawa find a splitting of 45\,meV between the two strongest peaks 
in the calculated RIXS response, in excellent agreement with the experimental data at 10\,K.\@ 
The fact that the splitting depends on $g$ reflects the dynamic Jahn-Teller effect,  
lifting the degeneracy of the $j$\,=\,3/2 quartet. 
This is in particular relevant for the temperature dependence of the splitting, as discussed 
in the next paragraph. Strictly speaking, the splitting also depends on $\lambda$ due to 
a finite admixture of $j$\,=\,1/2 character via the pseudo-Jahn-Teller effect 
\cite{Iwahara23,Liu19}. This, however, is small and can be neglected.

\subsection{Temperature dependence}

The convincing description of the temperature dependence of the RIXS data 
is a particular asset of the vibronic scenario.\@ Using $g$\,=\,1.2, 
Iwahara and Furukawa \cite{Iwahara23} showed that a vibronic model with spin-orbital-lattice 
entangled states describes the temperature dependence observed in RIXS.\@ 
First, we address the qualitative behavior of the experimental data, 
see Figs.\ \ref{fig:data_all}(a) and \ref{fig:rixs_fit}.
This can be understood within the Franck-Condon picture, see Eq.\ \eqref{eq:vib} 
and the black and orange lines in Fig.\ \ref{fig:FC}(b).
The thermal occupation of phonon modes gives rise to a redistribution of the intensity to 
high energies as well as to the range below the main peak, where the latter can be attributed 
to the emergence of a subband at $E_0-E_{\rm p}$.

As at 10\,K, a quantitative analysis reveals fingerprints of the entangled character.
With increasing temperature, the fit yields an increase of the peak splitting 
$E_{\rm p}$ and of $S$ and a modest red shift of $E_0$, see Fig.\ \ref{fig:rixs_fit}.
Remarkably, the behavior of all three observations is in line with an increase of the 
electron-phonon coupling constant $g$.
An increase of $g$ enhances $S$ and, in a spin-orbital-lattice entangled 
scenario \cite{Iwahara23}, reduces the energy $E_0$ of the main peak 
via the dynamic Jahn-Teller effect while increasing 
the splitting $E_{\rm p}\cdot (1+g^2/8)$ of the dominant RIXS peaks, as discussed above.
In fact, the fit yields a splitting as large as 51\,meV at 300\,K.\@ This is hard to 
reconcile with the phonon energies of the chloride in a Franck-Condon scenario 
but can be attributed to the $g$ dependence of the splitting in a spin-orbital-lattice 
entangled picture.

A further contribution to the 1\,\% reduction of $E_0$ may originate from thermal expansion 
of the lattice and a corresponding increase of the Ir-Cl distance. 
This modifies the cubic crystal field 10\,$Dq$ and thereby the effective value of $\lambda$. 
For the cubic phase of K$_2$IrBr$_6$, the peak splitting has been reported 
by Khan \textit{et al.}\ \cite{Khan21} to increase from about 50\,meV at 170\,K to 
70\,meV at 300\,K, while Reig-i-Plessis \textit{et al.}\ find a smaller value of 47\,meV 
at 300\,K \cite{Reig20}. 
This discrepancy may originate from the line shape of the spin-orbit exciton in 
K$_2$IrBr$_6$ which rather shows a subtle shoulder instead of a clear splitting.  
	
The temperature dependence of $\sigma_1(\omega)$ predominantly reflects the 
phonon-assisted character, which causes a pronounced increase of the spectral weight. 
The many subbands visible at 10\,K are smeared out at elevated temperatures, 
see Fig.\ \ref{fig:data_all}(b). 
Therefore we refrain from fitting $\sigma_1(\omega)$ at high temperature.

\subsection{Phonons and phonon sidebands}
\label{sec:phonon}

\begin{figure}[t]
	\centering
	\includegraphics[width=\columnwidth]{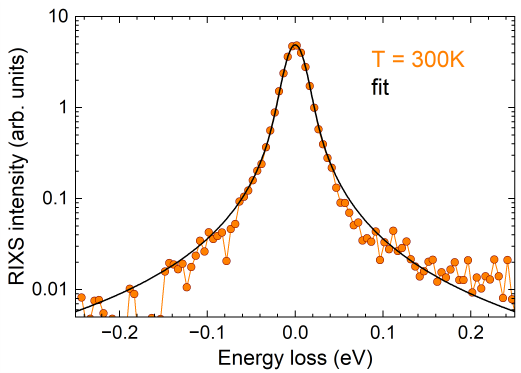}
	\caption{RIXS data of K$_2$IrCl$_6$ around zero energy loss at 300\,K.\@ The data are plotted 
		on a logarithmic scale to emphasize the tails of the elastic line that is fitted 
		using a Voigt line shape with the instrumental resolution $\delta E$\,=\,25\,meV.\@  
		The symmetric behavior around zero loss down to the noise level of about 0.01 
		provides no indication of low-energy excitations. In particular, the intensity 
		of phonons is negligible in RIXS at the Ir $L_3$ edge.}
	\label{fig:elastic}
\end{figure}

The observation of vibronic "phonon" sidebands is particularly interesting 
given the fact that the intensity of elementary phonon excitations is negligible in 
$L$-edge RIXS on $5d^5$ Ir$^{4+}$ $j$\,=\,1/2 Mott insulators \cite{Revelli20}. 
For K$_2$IrCl$_6$, this is highlighted in Fig.\ \ref{fig:elastic}, which focuses on the 
RIXS response around zero loss. The data have been measured with a scattering angle 
$2\theta$ close to 90$^\circ$, which yields an elastic line of moderate strength. 
The symmetric behavior around zero loss is emphasized by a fit with a Voigt profile. 
The data are very well described by considering only the elastic line and do not provide 
any evidence for a sizable inelastic contribution below 0.1\,eV.\@ 
Even at 300\,K, where the intensity of low-energy modes is enhanced by the Bose factor, 
the contribution of phonons is negligible.

In $L$-edge RIXS on Mott-insulating $5d^5$ iridates, spin and orbital excitations 
are boosted in a direct RIXS process \cite{Ament11}. 
The x-ray absorption step from the ground state to the intermediate state 
is followed by the x-ray emission step from the intermediate state to the final state. 
In general, phonons are excited with a much smaller cross section in 
\textit{indirect} RIXS, i.e., 
via the dynamics in the \textit{intermediate} state. In RIXS a phonon is created if 
the lattice distorts in the intermediate state to screen the core hole  
\cite{Ament11EPL,Geondzhian20,Gilmore23,Braicovich20,Devereaux16}.
A simplified description of phonon excitations can again be achieved 
in a Franck-Condon picture. However, with the coupling taking place in the intermediate state, 
one has to use \textit{two} Franck-Condon factors, one for the absorption step, 
and the second one for x-ray emission \cite{Ament11EPL,Geondzhian20,Gilmore23,Braicovich20}. 
For $L$-edge RIXS with an incident energy as large as 11.214\,keV, the absence of any 
phonon signatures can be rationalized via 
the Ir $2p^5$ $t_{2g}^6$ intermediate state, which is well screened 
and shows a short life time of only a few femtoseconds. 
In contrast, phonon contributions can be observed in RIXS on Ir oxides at 
the O $K$ edge due to the very different intermediate state \cite{Vale19,Ament11EPL,Geondzhian20,Gilmore23}.

Previously, this phonon approach with two Franck-Condon factors has also been 
applied to vibronic sidebands of on-site $d$-$d$ excitations studied in RIXS at the O $K$ edge, 
e.g., in Li$_2$CuO$_2$ and $\alpha$-Li$_2$IrO$_3$ \cite{Johnston15,Vale19}, as well as at the 
Cu $L$ edge in Ca$_2$Y$_2$Cu$_5$O$_{10}$ \cite{Lee14}. 
In our data of K$_2$IrCl$_6$, measured at the Ir $L_3$ edge, indirect RIXS processes are negligible.
The spin-orbit exciton and its vibronic sidebands are resonantly enhanced in a direct RIXS process 
due to the electronic contribution to the wave function. 
Empirically, the data are described by Eq.\ \eqref{eq:vib}, where the squared 
matrix element of the electronic transition in direct RIXS yields $I_0$. 
As in optical absorption, the electronic excitation 
leaves the system in a vibrationally excited state, see Fig.\ \ref{fig:FC}. 
In this sense, the coupling to the lattice occurs in the final state, in contrast 
to the case of elementary phonons discussed above.

\section{Double spin-orbit exciton}

\begin{figure}[t]
	\centering
	\includegraphics[width=\columnwidth]{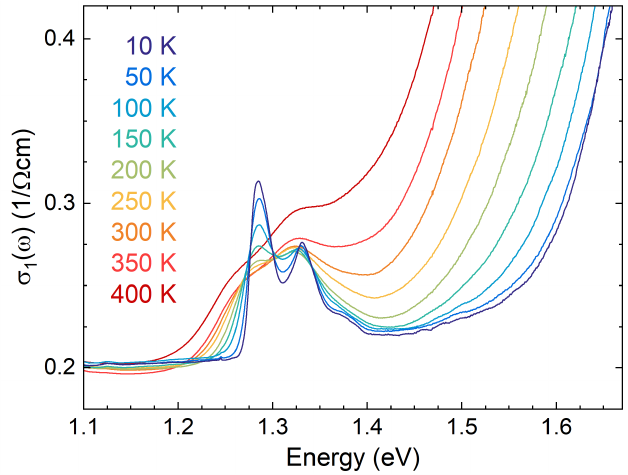}
	\caption{Double spin-orbit exciton around 1.3\,eV with vibronic sidebands. 
		Up to about 200\,K, the three-peak structure of this very weak absorption band 
		is located right below the onset of excitations across the Mott gap. }
	\label{fig:double}
\end{figure}

The excitation of double spin-orbit excitons in the optical conductivity around 1.3\,eV provides 
a further example for a vibronic excitation in K$_2$IrCl$_6$, see Fig.\ \ref{fig:double}. 
The feature is very weak, the spectral weight being roughly a factor 50 smaller than for the 
single spin-orbit exciton, see inset of Fig.\ \ref{fig:data_all}(b). 
At 10\,K, the energy of the first peak $E_\mathrm{2SO}$\,=\,1286\,meV\,=\,2$\,\cdot\,$643\,meV 
is very close to twice the energy $E_0$\,=\,635\,meV for the single spin-orbit exciton. 
This peak at 1.3\,eV can hence be assigned to the simultaneous excitation of two spin-orbit excitons on 
neighboring sites. Double and even triple spin-orbit excitons previously have been observed in 
$\sigma_1(\omega)$ of the $4d^5$ $j$\,=\,1/2 compound $\alpha$-RuCl$_3$ \cite{Warzanowski20,Lee21}.
Similar overtones or double intra-$t_{2g}$ excitations have also been reported in orbitally 
ordered YVO$_3$ as well as in K$_2$ReCl$_6$ and K$_2$OsCl$_6$ \cite{Bettinelli88,Warzanowski23,Warzanowski24,Benckiser08}.

The spectral weight of this feature is insensitive to temperature at least up to 200\,K.\@
At still higher temperature, it drowns in the onset of excitations across the Mott gap 
which shifts to lower energies with increasing temperature. 
The temperature-independent spectral weight suggests that this feature is \textit{not} phonon-assisted 
but rather directly infrared allowed, even though with a very small spectral weight. 
This agrees with results on K$_2$ReCl$_6$ and K$_2$OsCl$_6$ \cite{Warzanowski23,Warzanowski24}. 
A finite dipole moment may arise if the total excited state breaks inversion symmetry on the bond. 
This is, e.g., the case for a spin-orbit exciton with $j_z$\,=\,3/2 on one site but 
$j_z$\,=\,1/2 on the neighboring site. 
At 10\,K, the splitting between the main peak at $E_\mathrm{2SO}$\,=\,1286\,meV and the first 
sideband amounts to 46\,meV.\@ The splitting between the second peak and the weak third peak 
is comparable but the large width prevents a more precise determination. The value of 46\,meV 
is very similar to the vibronic splitting of the spin-orbit exciton, 
see Fig.\ \ref{fig:rixs_fit}, as well as to the 
vibronic splitting observed for the overtones in K$_2$ReCl$_6$ \cite{Warzanowski24}. 
The line shape, the experimental peak splitting, and the analogy with the sister compound 
K$_2$ReCl$_6$ provide strong evidence for a vibronic character of the double spin-orbit exciton.

\section{Discussion and Conclusion}

Spin-orbit entangled $j$\,=\,1/2 moments are Kramers doublets and as such not Jahn-Teller 
active. In frustrated quantum magnets, magneto-elastic coupling nevertheless may give rise 
to distortions that break cubic symmetry and lift magnetic frustration in $j$\,=\,1/2 compounds \cite{Revelli19}. 
A possibly pronounced magneto-elastic coupling would play an essential role for the low-energy 
physics of $j$\,=\,1/2 quantum magnets. 
We combined RIXS, optical spectroscopy, and single-crystal x-ray diffraction to study 
the possible role of magneto-elastic coupling in antifluorite-type K$_2$IrCl$_6$ 
with $j$\,=\,1/2 moments on an fcc lattice with highly frustrated exchange couplings. 
The global cubic structure of K$_2$IrCl$_6$ is well established \cite{Khan19,Reig20,Bertin24}. 
The sizable atomic displacement parameters reported for K$_2$IrCl$_6$ at elevated temperatures \cite{Khan19,Bertin24} 
exhibit a linear temperature dependence, underlining their dynamical origin based on 
rigid, low-energy rotations of the IrCl$_6$ octahedra. 
Even though the diffraction data do not provide any indication for deviations from cubic 
symmetry, the presence of local distortions cannot fully be excluded. 
Small distortions are a challenge for diffraction experiments and often are detected 
more easily in spectroscopy based on, e.g., a change of selection rules. 
The two-peak structure of the spin-orbit exciton observed in $L$-edge RIXS around 0.63\,eV 
can be attributed to either a noncubic crystal-field splitting or to a vibronic character 
with phonon sidebands. 
Optical spectroscopy provides complementary information based on the excellent energy 
resolution and the different excitation process with different selection rules. 
We observe a multitude of phonon-assisted peaks in $\sigma_1(\omega)$ that reach as high 
as 0.74 and 0.77\,eV.\@ These features require an explanation in terms of vibronic sidebands, 
in agreement with previous results on IrCl$_6$ impurities in host crystals \cite{Keiderling75,Yoo86}, 
with the vibronic character of the double spin-orbit exciton that we find around 1.3\,eV, and 
with optical results on related intra-$t_{2g}$ excitations in the sister compounds K$_2$ReCl$_6$ 
and K$_2$OsCl$_6$ \cite{Warzanowski23,Warzanowski24}. 
Furthermore, the vibronic picture is able to describe the temperature 
dependence observed in RIXS \cite{Iwahara23}. 
The success of the vibronic scenario suggests that magneto-elastic coupling is not decisive 
for the stability of the $j$\,=\,1/2 moments in K$_2$IrCl$_6$. 
Altogether, we conclude that K$_2$IrCl$_6$ hosts cubic $j$\,=\,1/2 moments 
and spin-orbital-lattice entangled $j$\,=\,3/2 excited states.

This spin-orbital-lattice entangled nature of the spin-orbit exciton is caused by 
electron-phonon coupling that yields a hybridization of the Jahn-Teller active $j$\,=\,3/2 
quartet with vibrational states \cite{Iwahara23}. 
The Franck-Condon approximation offers an analytic description of the	vibronic line shape. 
It assumes a negligible timescale for electronic excitations, approximating the eigenstates 
as product states of electronic and vibrational states. 
Empirically, the Franck-Condon approach describes the data of RIXS and optics well 
and in an intuitive way. 
Equivalent to the analysis of optical data, a single Franck-Condon factor 
is appropriate to describe vibronic excitations studied in \textit{direct} RIXS.\@ 
This differs from the case of elementary phonons that contribute in an indirect RIXS process, 
and we have shown that this contribution is negligible in our $L$-edge RIXS data of K$_2$IrCl$_6$. 
This intuitively highlights the particular character of the vibronic "phonon" sidebands of 
the spin-orbit exciton, for which resonance enhancement in direct RIXS applies to the 
electronic part of the wave function.

A more detailed quantitative description of the experimental data 
reveals the limitations of the Franck-Condon scenario, 
in which the splitting is equal to the phonon energy. 
In particular, the large splitting of 44\,meV observed at 10\,K and even 51\,meV at 300\,K 
is hard to reconcile with an elementary phonon energy in K$_2$IrCl$_6$. 
In contrast, a theory describing the microscopic coupling of the spin-orbit entangled states 
to the $e_g$ phonon mode with energy $E_{\rm p}$ yields $E_{\rm p}\cdot (1+g^2/8)$ 
for the splitting of the first two strong RIXS peaks \cite{Iwahara23}.
We thus conclude that the large splitting and its temperature dependence are fingerprints 
of the spin-orbital-lattice entangled character beyond the Franck-Condon approximation.

Concerning $L$-edge RIXS, the well resolved vibronic sideband in K$_2$IrCl$_6$ stands out 
in transition-metal compounds in general as well as compared to other $5d^5$ iridates. 
The splittings larger than 0.1\,eV observed in Ir$^{4+}$ oxides 
\cite{Liu12,Gretarsson13,Aczel19,Revelli19,Ruiz21,delaTorre21,Jin22,Magnaterra23Ti,delaTorre23,Kim12,Kim14,Lu18,Kim12327,Moretti15} 
are caused by a noncubic crystal field. Vibronic sidebands of electronic excitations 
have been claimed in Ca$_2$Y$_2$Cu$_5$O$_{10}$ \cite{Lee14}, 
K$_2$RuCl$_6$ \cite{Iwahara23Ru}, as well as in $5d^1$ compounds based on the asymmetric 
line shape of the excitation from $j$\,=\,3/2 to 1/2 \cite{Agrestini24,Frontini24}, 
but individual peaks have not been resolved. Typically, the subbands of vibronic 
excitations are not resolved in solids, even in optical data. 
However, the vibronic features are particularly sharp in $\sigma_1(\omega)$ of 
K$_2$ReCl$_6$ \cite{Warzanowski24} and still very well resolved in K$_2$IrCl$_6$, 
even in RIXS.\@ This can be attributed to the quasimolecular crystal structure 
with well separated $M$Cl$_6$ octahedra. 
The antifluorite-type $A_2MX_6$ compounds thus offer an ideal platform to further 
investigate the role of vibronic effects in RIXS.

\section*{Appendix}

\subsection{Crystal structure determination}

\begin{table}[t]
	\begin{tabular}{ccccccccc}
		\hline
		T/K  &  & $U_{11}$(Ir) && $U_{11}$(Cl) && $U_{22}$(Cl) && $U_{11}$(K) \\
		\hline
		290   & I  &	0.01584(8) && 0.0166(3)	&& 0.0425(3) && 0.0382(3) \\
		& II &	0.01557(7) && 0.0164(2)	&& 0.0422(2) && 0.0379(2) \\
		270  & I  & 0.01466(7) && 0.0153(2) && 0.0397(2) && 0.0353(2) \\
		& II & 0.01475(7)	&& 0.0156(2) && 0.0399(2) && 0.0355(2) \\
		250 & I	  & 0.01360(8)	&& 0.0138(3) && 0.0373(3) && 0.0327(3) \\
		& II  & 0.01354(7)	&& 0.0145(2) && 0.0369(2) && 0.0327(2) \\
		230	& I	  & 0.01219(6) && 0.01285(18) && 0.03407(17) && 0.02996(17) \\
		& II  & 0.01250(7)	&& 0.0132(2) && 0.0346(2) && 0.0302(2) \\
		210	& I	  & 0.01140(7) && 0.0120(2) && 0.0317(2) && 0.0277(2)  \\
		& II  & 0.01157(7) && 0.0121(2) && 0.0320(2) && 0.0279(2) \\
		190	& I   & 0.01035(6) && 0.0110(2) && 0.02898(19) && 0.02500(19) \\
		& II  & 0.01045(7) && 0.0110(2) && 0.0290(2) && 0.0250(2) \\
		170	& I   & 0.00930(6) && 0.00981(19) && 0.02660(18) && 0.02249(17) \\
		& II  & 0.00931(7) && 0.0098(2) && 0.0263(2) && 0.0226(2) \\
		150	& I   & 0.00836(6) && 0.0090(2) && 0.02420(18) && 0.02019(18) \\
		& II  & 0.00793(5) && 0.00871(16) && 0.02343(14) && 0.01984(14) \\
		130 & I	  & 0.00759(12) && 0.0082(3) && 0.0216(3) && 0.0180(3) \\
		& II  & 0.00729(6) && 0.00797(19) && 0.02123(17) && 0.01761(16) \\
		120	& I	  & 0.00686(6) && 0.00739(19) && 0.02010(17) && 0.01635(16) \\ 
		\hline        
	\end{tabular}
	\caption{Anisotropic atomic displacement parameters $U_{ij}$, derived from crystal structure determination by single crystal x-ray diffraction. The data were measured in the sequence from 290 to 120\,K (I) and back to 290\,K (II). 
	Figure \ref{fig:ADP} shows the average of I and II.\@ 
	All $U_{ij}$ are given in units of \AA$^2$.}
\end{table}

The x-ray diffraction experiments for a crystal of K$_2$IrCl$_6$ with dimensions of 
0.072\,$\times$\,0.059\,$\times$\,0.049\,mm$^3$ were performed on a Bruker D8 VENTURE Kappa Duo PHOTONIII diffractometer with a I$\mu$S micro-focus sealed tube (Mo K$_\alpha$ radiation, 0.71073\,\AA). Using an Oxford Cryostream Cooler800, 
we collected data between 290 and 120\,K.\@ The crystal structure was solved by direct methods (SHELXT) \cite{Sheldrick15A} and refined by full matrix least squares based on $F^2$ (SHELXL2019) \cite{Sheldrick15C}. 
Multi-scan absorption correction was applied. The anisotropic atomic displacement parameters $U_{ij}$ 
are given in Table I.\@ The data were obtained by first cooling down in steps from 290 to 120\,K, 
while a second set of data points was measured upon heating back to 290\,K.\@

The x-ray crystallographic data have been deposited at the Inorganic Crystal Structure Database via the joint CCDC/FIZ Karlsruhe deposition service \cite{Database} with the deposition numbers CCDC/CSD 2367387 to 2367405 and can be obtained free of charge from https://www.ccdc.cam.ac.uk/structures.

\begin{figure}[b]
	\centering
	\includegraphics[width=0.95\columnwidth]{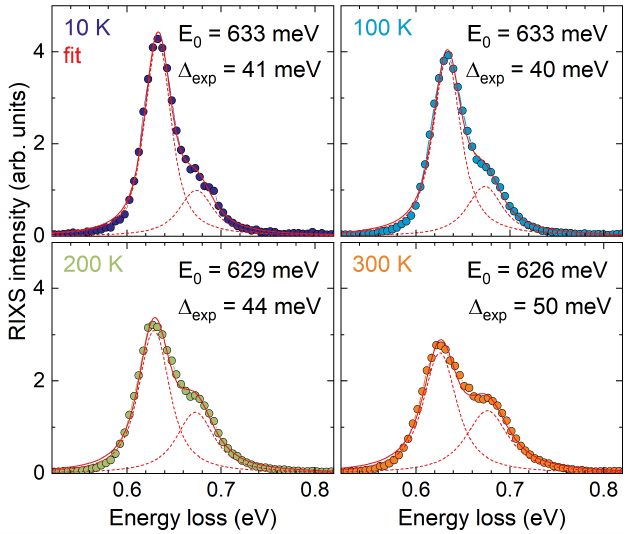}
	\caption{Fits of the RIXS data using two Voigt profiles. 
		The two features (dashed) are peaking at $E_0$ and $E_0 + \Delta_{\rm exp}$. 
		The solid red line depicts the sum.  }
	\label{fig:rixs_fit_voigt}
\end{figure}

\subsection{Two-peak fit of RIXS}

Phenomenologically, the apparent two-peak structure of the RIXS data can also be 
described by employing two oscillators with a Voigt line shape, 
see Fig.\ \ref{fig:rixs_fit_voigt}. The Voigt profile corresponds to a convolution 
of Gaussian and Lorentzian peaks. For the Gaussian part we use the instrumental width 
determined from the elastic line. These fits yield the energy $E_0$ of the dominant peak 
and the peak splitting $\Delta_{\rm exp}$ of 40 to 50 meV.\@ 
Compared to the vibronic fit shown in Fig.\ \ref{fig:rixs_fit}, we find the same result 
for $E_0$ while the peak splitting $\Delta_{\rm exp}$ is slightly smaller than the 
phonon progression energy $E_p$. 
The agreement between the two-peak Voigt fits and the experimental data does 
not achieve the same quality as for the vibronic fits shown in Fig.\ \ref{fig:rixs_fit}. 
This applies in particular to the low-frequency side below 0.6\,eV 
and reflects the extended tails of the Lorentzian contribution.

\subsection{Optical conductivity above the Mott gap}

\begin{figure}[t]
	\centering
	\includegraphics[width=\columnwidth]{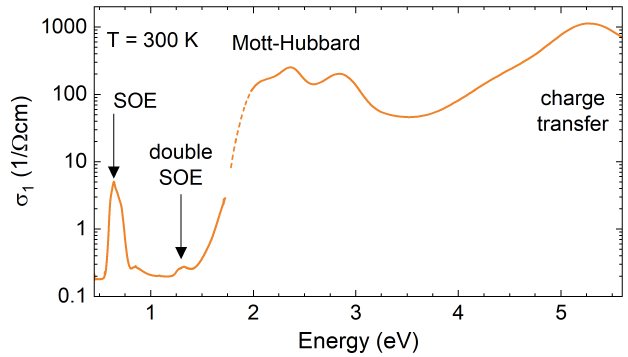}
	\caption{Optical conductivity up to 5.6\,eV at room temperature. 
		Data above the Mott gap have been determined by ellipsometry measurements. 
		Note the logarithmic scale. Above the Mott gap, $\sigma_1(\omega)$ is roughly 
		two orders of magnitude larger than for the spin-orbit exciton.  
		The dashed line depicts a fit that assumes a Tauc-Lorentz shape of the Mott gap. }
	\label{fig:elli}
\end{figure}

We address the optical conductivity at higher energies to put the phonon-assisted 
features into context. 
Figure \ref{fig:elli} shows $\sigma_1(\omega)$ above the Mott gap as determined 
by ellipsometry. Above about 4\,eV, we observe charge-transfer excitations between 
Ir and Cl sites. 
In the energy range from 2 to 3.5\,eV we find Mott-Hubbard excitations across the Mott gap, 
i.e., \textit{intersite} excitations $|d_i^5 d_j^5\rangle \rightarrow |d_i^4 d_j^6\rangle$ 
between the Ir sites $i$ and $j$. Compared to \textit{on-site} $d$-$d$ excitations such as the 
spin-orbit exciton, the value of $\sigma_1(\omega)$ for these directly dipole-allowed 
excitations is about two orders of magnitude larger. Accordingly, the strong intersite 
excitations cover the on-site crystal-field excitations from $t_{2g}$ to $e_g$ states. 
These are peaking in the energy range from 2.5 to 3.5\,eV in RIXS on K$_2$IrCl$_6$ 
\cite{Reig20} but are not visible in the optical data. 
Considering the additional energy cost of the on-site Coulomb repulsion $U$ for 
intersite excitations across the Mott gap, the two lowest peaks at about 2.4 and 2.9\,eV 
have to be attributed to $t_{2g}$ states. Their energies can be rationalized in terms 
of local multiplets. This approach is corroborated by the observation that the 
spectral weight of such intersite excitations in many transition-metal compounds 
reflects spin and orbital correlations between nearest neighbors 
\cite{Khaliullin04,Oles05,Fang03,Kovaleva04,Lee05,Goessling08,Reul12,Vergara22}. 
In this picture, the lowest $|d_i^4 d_j^6\rangle$ excited states 
exhibit a $t_{2g}^6$ $^1A_1$ multiplet on site $j$ while site $i$ hosts a 
$t_{2g}^4$ multiplet with either $^3T_1$ or $^1T_2/^1E$ symmetry. 
In a $t_{2g}$-only picture, the corresponding excitation energies 
$E(t_{2g}^4)+E(t_{2g}^6)-2E(t_{2g}^5)$ amount to $U-3J_H$ and $U-J_H$ \cite{Zhang17}. 
The energy difference of 2\,$J_H$ is reduced to about 1.5\,$J_H$ upon taking into 
account the interaction with $e_g$ orbitals \cite{Warzanowski23}. 
The observed energy difference of 0.5\,eV thus agrees with the expectation 
$J_H$\,$\approx$\,0.3-0.4\,eV \cite{Warzanowski23,KKM14,Winter17}.

Finally, we want to point out the close relation to intersite excitations in $d^1$ 
compounds such as Mott insulating YTiO$_3$ \cite{Goessling08}. There, the energy of 
the $|d_i^0 d_j^2\rangle$ excited states is governed by two-electron $t_{2g}^2$ 
multiplets that show the same symmetry as the two-hole $t_{2g}^4$ states discussed 
above for K$_2$IrCl$_6$. 
Remarkably, the lowest intersite excitation in YTiO$_3$, involving the $^3T_1$ multiplet, 
exhibits a two-peak structure that has been explained in terms of a Mott-Hubbard 
exciton \cite{Goessling08} for which the energy is reduced by the Coulomb interaction 
of an electron-hole pair on nearest-neighbor sites as well as by magnetic and orbital 
correlations. The term Mott-Hubbard exciton is employed not only for truly bound excitons 
below the Mott gap but also for a nearly-bound resonance within the continuum, 
as observed in YTiO$_3$. 
Mott-Hubbard excitons have been discussed also for the $5d^5$ iridates 
Na$_2$IrO$_3$ and Sr$_2$IrO$_4$ \cite{Alpichshev15,Mehio23}.
In this light, one may speculate that the pronounced shoulder at 2\,eV of the peak at 
2.5\,eV in $\sigma_1(\omega)$ of K$_2$IrCl$_6$ also reflects nearest-neighbor Coulomb 
interactions.

\begin{acknowledgments}
	We thank K. Hopfer and H. Schwab for experimental support and the European Synchrotron Radiation Facility for providing beam time at ID20 and technical support. 
	Furthermore, we acknowledge funding from the Deutsche Forschungsgemeinschaft 
	(DFG, German Research Foundation) through Project No.\ 277146847 -- CRC 1238 
	(projects A02, B02, B03) 
	as well as from the European Union – Next Generation EU - “PNRR - M4C2, investimento 1.1 - Fondo PRIN 2022” - “Superlattices of relativistic oxides” (ID 2022L28H97, CUP D53D23002260006).
\end{acknowledgments}

\end{document}